\documentclass{aa}  

\usepackage{graphicx}
\usepackage{txfonts}
\newcommand{\hii}{\relax \ifmmode {\mbox H\,{\scshape ii}}\else H\,{\scshape ii}\fi}
\newcommand{\mi}{\relax \ifmmode {\mu{\mbox m}}\else $\mu$m\fi}
\newcommand{\ha}{\relax \ifmmode {\mbox H}\alpha\else H$\alpha$\fi}
\newcommand{\hb}{\relax \ifmmode {\mbox H}\beta\else H$\beta$\fi}

\newcommand{\sii}{\relax \ifmmode {\mbox S\,{\scshape ii}}\else S\,{\scshape ii}\fi}
\newcommand{\siii}{\relax \ifmmode {\mbox S\,{\scshape iii}}\else S\,{\scshape iii}\fi}
\newcommand{\nii}{\relax \ifmmode {\mbox N\,{\scshape ii}}\else N\,{\scshape ii}\fi}
\newcommand{\oi}{\relax \ifmmode {\mbox O\,{\scshape i}}\else O\,{\scshape i}\fi}
\newcommand{\oii}{\relax \ifmmode {\mbox O\,{\scshape ii}}\else O\,{\scshape ii}\fi}
\newcommand{\hei}{\relax \ifmmode {\mbox He\,{\scshape i}}\else He\,{\scshape i}\fi}
\newcommand{\heii}{\relax \ifmmode {\mbox He\,{\scshape ii}}\else He\,{\scshape ii}\fi}
\newcommand{\oiii}{\relax \ifmmode {\mbox O\,{\scshape iii}}\else O\,{\scshape iii}\fi}
\newcommand{\ariii}{\relax \ifmmode {\mbox Ar\,{\scshape iii}}\else Ar\,{\scshape iii}\fi}
\newcommand{\ariv}{\relax \ifmmode {\mbox Ar\,{\scshape iv}}\else Ar\,{\scshape iv}\fi}
\newcommand{\neiii}{\relax \ifmmode {\mbox Ne\,{\scshape iii}}\else Ne\,{\scshape iii}\fi}

\newcommand{\rdostres}{\relax \ifmmode {\,\mbox{R}}_{\rm 23}\else \,\mbox{R}$_{\rm 23}$\fi} 

\newcommand{\ciii}{\relax \ifmmode {\mbox O\,{\scshape iii}}\else C\,{\scshape iii}\fi}
\newcommand{\civ}{\relax \ifmmode {\mbox O\,{\scshape iii}}\else C\,{\scshape iv}\fi}

\newcommand{\gsim}{\hbox{\rlap{\lower.55ex\hbox{$\sim$}} \kern-.3em
\raise.4ex \hbox{$>$}}}
\newcommand{\lsim}{\hbox{\rlap{\lower.55ex\hbox{$\sim$}} \kern-.3em
\raise.4ex \hbox{$<$}}}

\usepackage{color}

\begin{document}

   \title{The softness diagram for MaNGA star-forming regions: diffuse ionized gas contamination or local HOLMES predominance?}

\titlerunning{The softness diagram in MaNGA}

\author{
        E.~P\'{e}rez-Montero\inst{\ref{IAA}}  \and
        I.~A.~Zinchenko\inst{\ref{LMU},\ref{MAO}} \and
        J.~M. V\'{i}lchez\inst{\ref{IAA}}  \and
        A.~Zurita\inst{\ref{UGR},\ref{ici}} \and 
        E.~Florido\inst{\ref{UGR},\ref{ici}} \and 
        B.~P\'{e}rez-D\'{i}az\inst{\ref{IAA}}
}
       
\institute{
Instituto de Astrof\'{i}sica de Andaluc\'{i}a (CSIC), Apartado 3004, 18080 Granada, Spain \label{IAA} 
\and
Faculty of Physics, Ludwig-Maximilians-Universit\"{a}t, Scheinerstr. 1, 81679 Munich, Germany \label{LMU}
\and
Main Astronomical Observatory, National Academy of Sciences of Ukraine, 
27 Akademika Zabolotnoho St., 03680, Kyiv, Ukraine\label{MAO}
\and
Dpto. de F\'\i sica Te\'orica y del Cosmos, Campus de Fuentenueva, Edificio Mecenas, Universidad de Granada, E18071--Granada, Spain\label{UGR}
\and
Instituto Carlos I de F\'isica Te\'orica y Computacional, Facultad de Ciencias, Universidad de Granada, E18071--Granada, Spain\label{ici}
}

   \date{Received XXX; accepted YYY}

 
  \abstract
   {} 
 {We explore the so-called {\em softness diagram} ---whose main function is to provide the hardness of the ionizing radiation in
        star-forming regions--- in order to check whether hot and old low-mass evolved stars (HOLMES) are significant contributors 
        to the ionization within star-forming regions, as suggested by previous MaNGA data analyses.} 
{We used the code {\sc HCm-Teff} to derive both the ionization parameter and the equivalent effective temperature ($T_*$), adopting models of massive stars and planetary nebulae (PNe), and exploring different sets of emission lines in the softness diagram to figure out the main causes of
        the observed differences in the {\em softness parameter} in the MaNGA and CHAOS star-forming region samples.}
        {
We find that the fraction of  regions with a resulting $T_* > 60 kK$, which are supposedly ionised by sources harder than massive stars, is considerably larger in the MaNGA (66\%) than in the CHAOS (20\%) sample when
the [\sii] $\lambda\lambda$ 6716,6731 \AA\ emission lines are used in the softness diagram.
However, the respective fractions of regions in this regime for both samples are considerably reduced (20\% in MaNGA and 10\% in CHAOS) when the [\nii] emission line at $\lambda$ 6584 \AA\ is used instead.
This may indicate that diffuse ionised gas (DIG) contamination in the lower resolution MaNGA data is responsible for artificially increasing the measured $T_*$
 as opposed to there being a predominant role of HOLMES in the \hii\ regions.}
  {}


   \maketitle
%

\section{Introduction}
Optical emission lines have traditionally been used as bright tracers of the properties of the \hii\ regions where they are produced.
Among them, collisionally excited lines (CELs), given that they are among the brightest in the optical spectrum,
can be calibrated, either empirically or based on models, in order to derive several of the 
so-called functional parameters that govern the basic physics of the ionized gas, which include the metal content, the excitation,
or the hardness of the ionizing source.

Among the different strategies used to comparatively quantify the shape of the incident
spectral energy distribution (SED), \cite{vp88} defined the so-called {\em softness} parameter as

\begin{equation}
\eta = \frac{{\rm O}^+/{\rm O}^{2+}}{{\rm S}^+/{\rm S}^{2+}}
\end{equation}

\noindent       based on the assumption that the quotient of the two defined ionic abundance ratios minimizes the dependence on ionization parameter (i.e., the ratio between the number of hydrogen-ionizing photons and the density of particles, usually denoted as $U$) and
can be derived using only available optical CELs. The parameter is defined in such a way that lower values of $\eta$ correspond to harder ionizing SEDs. An alternative version of this parameter was also defined that  uses corresponding emission line fluxes, though a small dependence on the electron temperature is present in this case:

\begin{equation}
\eta\prime = \frac{[{\rm OII}] \lambda3727/[{\rm OIII}] \lambda\lambda4959,5007 }{[{\rm SII}] \lambda\lambda6717,6731 / [{\rm SIII}] \lambda\lambda9059,9532}
.\end{equation}

A different model-based approach was proposed by \cite{pm19}, given that using only the involved emission-line ratios  in a two-dimensional plot known as the {\em softness diagram} helps to separate the dependence of these ratios on metallicity, log $U,$ and the hardness of the incident SED.
In addition, when comparing the observed ratios with the results from photoionization models with a SED scalable in terms of an equivalent effective
temperature (e.g., massive single stars or blackbody), the diagram can provide a quantification of this hardness.

Although this methodology can be successfully used in certain \hii\ regions in disk galaxies, it does
not allow proper coverage of the space of observed emission line fluxes in certain regions
observed with IFS, such as in MaNGA \citep{kumari21}, as
this presents lower $\eta\prime$ values,  on average, and therefore a harder incident field of radiation.

Several authors (e.g., \citealt{mannucci21}) have recently studied    the average differences observed
between some of the emission-line ratios involved in $\eta\prime$, such as [\sii]/[\siii], in samples of \hii\ regions observed using long-slit spectrophotometry, such as CHAOS \citep{chaos-n628}, in
relation to other samples based on IFS, such as MaNGA.
According to these authors, the lower observed [\sii]/[\siii] ratios in CHAOS could be a consequence of
incomplete  coverage of the ionization structure, but
this is difficult to reconcile with the
fact that the [\oii]/[\oiii] is higher in the same regions as compared to IFS data and that photoionization models, which provide integrated emission-line fluxes, do not predict [\sii] fluxes as high as those observed
when only massive stars are considered.

As opposed to the study of a single low- to high-ionization emission-line flux ratio, the analysis of the softness diagram, could in turn point to the
predominance of very hard ionizing sources, such as the hot low-mass evolved stars (HOLMES; \citealp{flores11}).
The hard SED from HOLMES can significantly contribute to the [\oiii] and [\sii] emission (e.g., \citealt{belfiore22}) and, as shown by \cite{kumari21}, models with an older stellar population
like those calculated in \cite{morisset16} could in principle cover the region with very low $\eta\prime$ values observed for star-forming
         regions detected in MANGA data.

The application of a method such as {\sc HCm-Teff}, a bayesian-like code that can provide solutions comparing the predictions from models with the observational positions 
of data in the softness diagram, can shed light on this question by incorporating SEDs compatible with HOLMES and allowing the study of alternative emission lines in order to explore the observed differences between long-slit- and IFS-based samples.

In this work, we compiled emission-line data from both MaNGA and CHAOS, studying their position in the  softness diagram, and comparing them
with the results from models and the corresponding calculations made by {\sc HCm-Teff}. This study will allow us to disentangle the possible role of HOLMES and/or the contribution of the diffuse ionized gas (DIG) in the observations, quantifying the corresponding effect in the results.

The paper is organized as follows.
In Section 2 we describe the observational samples of compiled optical emission lines. In Section 3 we provide details of the computed photoionization models to be compared with the observations, and of the last version of the code {\sc HCm-Teff}.
In Section 4 we present our results and then discuss them, and finally in Section 5 we summarize our findings and provide conclusions. 

\section{Data samples}

\subsection{The MaNGA sample}

The Mapping Nearby Galaxies at Apache Point Observatory \citep[MaNGA;][]{Bundy2015} is part of the Sloan Digital Sky Survey IV \citep[SDSS IV;][]{Blanton2017}. For this work, we used data release 17 (DR17, \citealt{sdss-dr17}), from which we selected 2124 galaxies in the redshift range 0.0005 $< z <$ 0.0845.
We use individual spaxels whose size is significantly smaller than that  of the point spread function (PSF) in the MaNGA datacubes. The spatial resolution of these
cubes has a median  full width at half maximum (FWHM) of 2.54 arcsec \citep{Law2016}. 
As our sample has a median $z$ of 0.024, this corresponds to an average 
spatial resolution of 1.2 kpc.

We  collected emission-line fluxes of [\oii] $\lambda$3727 \AA, [\oiii] $\lambda\lambda$4959,5007 \AA, [\nii] $\lambda$6584 \AA, [\sii] $\lambda\lambda$6717,6731 \AA, [\ariii] $\lambda$7135 \AA, and [\siii] $\lambda\lambda$9069,9532 \AA, and obtained a signal-to-noise ratio (S/N) of at least 10 in all selected lines. This implies a total of 201 735 spectra. To obtain the emission line fluxes, we used the {\sc STARLIGHT} code \citep{CidFernandes2005,Mateus2006,Asari2007} to subtract the stellar background and the {\sc ELF3D} code to fit the emission lines. Details about the processing can be found in \citet{Zinchenko2016,z21}.  
The line fluxes were corrected for interstellar reddening using the analytical
approximation of the Whitford interstellar reddening law \citep{Izotov1994},
assuming the Balmer line ratio of $\text{H}\alpha/\text{H}\beta = 2.86$.
When the measured value of $\text{H}\alpha/\text{H}\beta$ is less than 2.86,
the reddening is set to zero.

\subsection{The CHAOS sample}

The CHAOS (CHemical Abundances Of Spirals) project has been collecting data 
to build a large database of high-quality, 1'' width, long-slit optical spectra of the   \hii\ regions of nearby spiral galaxies using the Multi-Object Double Spectrographs (MODS) on the Large Binocular Telescope (LBT).

We compiled data from the literature corresponding to the relative reddening-corrected emission-line fluxes necessary to build the various versions
of the softness diagrams studied in this work. In particular, we used data for NGC~628 \citep{chaos-n628}, NGC~2403 \citep{chaos-n2403},
NGC~3184 \citep{chaos-n3184}, NGC~5194 \citep{chaos-n5194}, and NGC~5457 \citep{chaos-n5457}. 
We used the relative to hb\ emission line fluxes of
these lines as reddening corrected by the corresponding authors,
but we checked that these corrections lead to results that are compatible, within the errors, with those obtained with the same fluxes corrected for reddening with the assumptions used for the MaNGA data previously described.

\begin{figure}
   \centering
\includegraphics[width=0.4\textwidth,clip=]{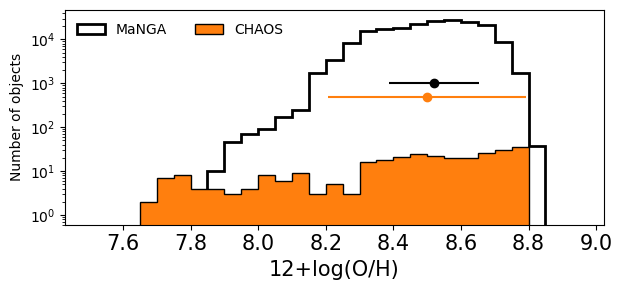}

\caption{Distribution of the total oxygen abundances derived for the two studied samples as obtained using {\sc HCm}.}

              \label{hist_oh}%
    \end{figure}

\section{Description of the models and the {\sc HCm-Teff} code}

We used photoionization models to interpret the position of the observed regions in the
softness diagram, and to provide predictions of the emission-line fluxes under different assumed conditions in order to derive the equivalent effective temperature ($T_*$) and the ionization parameter ($U$) with the {\sc Hii-Chi-mistry-Teff} code ({\sc HCm-Teff}\footnote{All versions of {\sc HCm} can be retrieved in the public site {\url http://www.iaa.csic.es/~epm/HII-CHI-mistry.html}}.
All photoionization models used in this work were calculated using the code {\sc Cloudy} v. 17 \citep{cloudy} and
are described in \cite{pm19}.
In addition to the SEDs described in this latter work, corresponding to O-stars from WM-Basic \citep{wmbasic} in the $T_*$ range from 30-60 kK, we 
incorporated SEDs from central stars of planetary nebulae (PNe) from \cite{rauch} in the range 80-120 kK to cover the $T_*$ range that is supposed to be reached by HOLMES.
All models were calculated assuming both spherical and plane-parallel geometry, also covering the range of log $U$ from -4.0 to -1.5, and for 
metallicities as scaled from the total oxygen abundance 12+log(O/H) from 7.1 and 8.9. The remaining chemical species assumed as input in the models were scaled to the solar proportions as given by \cite{asplund09}, except for N, which was scaled to the
empirical law given in \cite{pm14} (i.e., a constant N/O owing to a mostly primary production of N, and increasing N/O at larger Z when
secondary N production begins to be important). All models also consider a standard dust-to-gas mass ratio and a filling factor of 0.1. 

The {\sc HCm-Teff} code was presented and described in \cite{pm19}.
It can be  used to derive both $T_*$ and log $U$ by performing a bayesian-like comparison between the predictions from the
grids of models and some specific  emission-line flux ratios, involving reddening-corrected high and low-excitation emission-lines
(e.g., [\oii]/[\oiii], [\sii]/[\siii]).

It is known that the ratios of low- to high-excitation lines can be affected by the geometry of the gas if a fraction of ionizing photons
escape from the nebula. The code can also be used to derive this fraction using alternative grids of matter-bounded geometry models with {\sc Bpass} \citep{bpass} cluster atmospheres at different
stellar metallicities  \citep{pm20}.
However, for this work we only consider models assuming a radiation-bounded geometry.

Briefly, the code calculates $T_*$ and log $U$ as the average of the resulting $\chi^2$-weighted distributions of all the results of the models of the grids, with $\chi^2$ values being calculated as the quadratic differences between the observed and predicted corresponding emission-line ratios. The code also provides errors on the derived values, which are calculated as the quadratic addition of the standard deviations of the same distributions and the dispersion resulting from a Monte Carlo iteration perturbing the nominal emission-line flux values with the provided observational errors.

As shown by different authors (e.g., \citealt{morisset04,pmv09}), metallicity plays an important role in the position of the regions in the softness diagram, and so the code firstly interpolates the grids 
to the O/H value given as input in each region. 
If no O/H is given, the code finds a solution using the grid of models for all O/H values, which subsequently leads to a larger uncertainty, as discussed in \cite{pm19}. The error in the input oxygen abundance is also taken
by the code for its consideration in the error of the results in the Monte Carlo iteration.

Then, in order to better constrain the solutions for $T_*$ and $U$, we derived the oxygen abundance values in both samples  using {\sc HCm} version 5.22 \citep{pm14} and the models calculated with {\sc Popstar} \citep{popstar} SEDs,
and using as input all the collected emission lines mentioned above. Also, to this end we also used the auroral [\oiii] $\lambda$ 4363 \AA\ line in the case of a subsample of CHAOS (i.e., 134 \hii\ regions).
As discussed in several works \citep{pm14,pm16}, this method is totally consistent with the direct method, even in the absence of any auroral line,
although with a larger uncertainty (e.g., the mean O/H error in CHAOS for the \hii\ regions with [\oiii] $\lambda$ 4363 \AA\ is 0.11 dex, while for the rest it is 0.18 dex).
When no auroral line is given as an input, the code considers the empirical relation between O/H and excitation. In a similar way, when N/O cannot be directly calculated, the code assumes
an empirical law between O/H and N/O in order to derive metallicities.
For these  cases, we used the laws derived by \cite{pm14} for star-forming regions.

In the case of MaNGA, we did not use [\sii] lines as input for {\sc HCm} in order to derive chemical abundances, as the N2S2 ratio leads to very low values of N/O as compared to
those obtained with N2O2, as discussed by \cite{z21}. In the case of CHAOS, a certain contamination from iron emission to the [\oiii] $\lambda$ 4363 \AA\ has been reported in some galaxies of this sample
\citep{chaos-n2403}, but we checked that the effect of this difference in the final obtained oxygen abundances is not larger than the obtained errors and, in any case, does not lead
to noticeably different results for our derived $T_*$ and $U$.
In Figure \ref{hist_oh}, we show the distribution of the obtained oxygen abundances in both samples. Although MaNGA spread over larger metallicities, which is consistent with the fact that this sample includes more galaxies, covering a wider range of properties, both distributions present  an identical
median value of 8.52 ($\pm$ 0.13 for MaNGA, and $\pm$ 0.25 for CHAOS), which is around 0.7$\cdot$$Z_\odot$.

\begin{figure*}
   \centering
\includegraphics[width=0.45\textwidth,clip=]{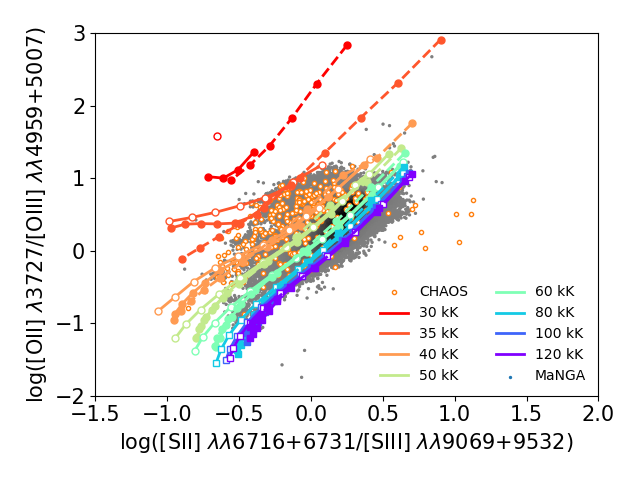}
\includegraphics[width=0.45\textwidth,clip=]{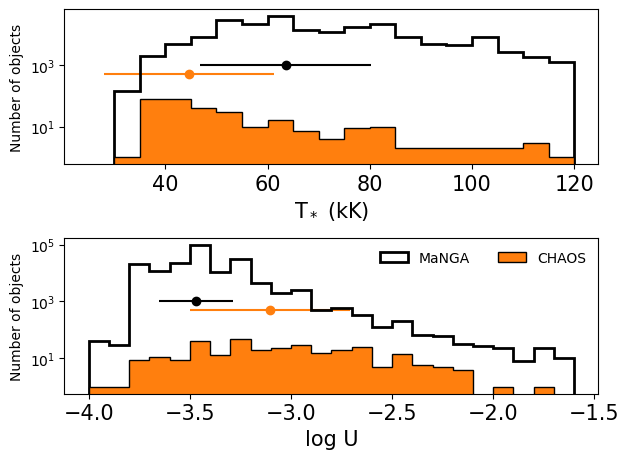}

\caption{Softness diagram and effective temperature and ionization parameter distributions in our samples. Left panel: Softness diagram representing [\oii]/[\oiii] vs [\sii]/[\siii] for the selected regions in CHAOS and  MaNGA as compared with different model sequences calculated using WM-Basic atmospheres for $T_*$ between 30 and 60 kK (represented with circles), and Rauch for $T_*$ between 80 and 120 kK (represented with squares). Solid lines represent models with 12+log(O/H) = 8.9, and dashed lines 12+log(O/H) = 8.0. Filled symbols represent models calculated assuming spherical geometry while empty symbols represent models with plane-parallel geometry (only for O/H = 8.9 in this case).
Right panel: Distributions of the derived properties using {\sc HCm-Teff} (upper panel: $T_*$, lower panel: log U) using as input the metallicity and the emission line-ratios represented in the left panel. The empty histograms represent the MaNGA sample and the orange ones the CHAOS sample. The median and standard deviations are represented in all panels in the same colors.}

              \label{o2o3-s2s3}%
    \end{figure*}

\section{Results and discussion}

\subsection{Softness diagram based on [\oii]/[\oiii] and [\sii]/[\siii]}

The left panel of Figure \ref{o2o3-s2s3} shows the softness diagram using the emission-line ratios [\oii]/[\oiii] versus [\sii]/[\siii] for
the selected regions in CHAOS and MaNGA as compared with the results from photoionization models with spherical geometry at two different metallicity values (i.e., 12+log(O/H) = 8.0 and 8.9). The same figure also shows some sequences of models calculated assuming plane-parallel geometry at the higher O/H value, which are used here to explore the effect of the assumed geometry on the results.
The represented models have  $T_*$ values in the range of 30-60 kK (i.e., using WM-Basic O star atmospheres) and 80-120 kK (using planetary nebulae (PNe) non-local thermodynamic equilibrium (non-LTE) stellar atmospheres), and they also cover a wide range in log $U$ going from
-4.0 to -1.5, which in the plot corresponds to the sequences going from the upper-right part to the lower-left regions of the plot, respectively.

We note that an important fraction of the two samples lies in a region of the diagram that could only be explained by invoking effective temperatures higher than the maximum value
reachable by O stars (i.e., 60 kK). Moreover, a fraction of the regions could have $T_*$ values larger than the maximum predicted by PNe even for the largest possible metallicity assumed in the models.
Although the fraction of points covered by the models with massive stars is slightly higher for spherical than for plane-parallel geometry, the discrepancy for most of the sample not covered by these models  cannot be attributed to the assumed geometry. Nonetheless, for the sake of simplicity, in what follows we only focus on the
results from models assuming spherical geometry, as this assumption is more realistic. In any case, this fraction of regions that would require very high $T_*$ according to this softness diagram is considerably smaller for CHAOS than for MaNGA.

In order to quantify this behavior, we used the code {\sc HCm-Teff} for both the selected MaNGA and CHAOS samples to derive $T_*$ and U.
The code accepts different emission lines, but we only introduced those involved in the softness diagram
shown in the left panel of Figure \ref{o2o3-s2s3}  (i.e., [\oii], [\oiii], [\sii] and [\siii])  as input
in a first step.
As an additional input, we introduced the derived metallicity for each point.
Although only two values of metallicity are represented in left panel of Figure \ref{o2o3-s2s3}, all possible values are considered by the code in order to cover the entire range and to properly interpolate to  the corresponding derived O/H in each point.

The right panel of Figure \ref{o2o3-s2s3}  shows  the distributions of the obtained values in both samples when a spherical geometry is assumed. When a plane-parallel geometry is assumed,
      the derived $T_*$ and log $U$ values are, on average, higher by 2kK and 0.01 dex, respectively.
These discrepancies are lower than the grid step for the corresponding values, meaning that the geometry assumed in the model has, on average, no significant impact on our results. 

Consistently with the position of the regions in both samples in the softness diagram, MaNGA presents a larger median $T_*$ (64 $\pm$ 17 kK) than in CHAOS (45 $\pm$ 16 kK).
The mean associated error for each derivation is around 9 kK for MaNGA and 5 kK for CHAOS. This difference is due to the fact that the resolution of the grid of models is lower for higher $T_*$. 
In fact, a significantly larger fraction of the regions in MaNGA could have $T_*$ $>$ 60 kK
than in CHAOS (66\% and 20\%, respectively).

A similar result is obtained for the distribution of derived log $U$ values, which is shown in the lower right panel of Figure \ref{o2o3-s2s3}, with a median value significantly lower for MaNGA (-3.47 $\pm$ 0.18) than for CHAOS (-3.10 $\pm$ 0.40),
in agreement with the fact that [\sii]/[\siii] is significantly higher in the first sample.
The typical error in the derivation of log$U$ is 0.06 dex for MaNGA and 0.10 dex for CHAOS. 

\begin{figure}
   \centering
\includegraphics[width=0.4\textwidth,clip=]{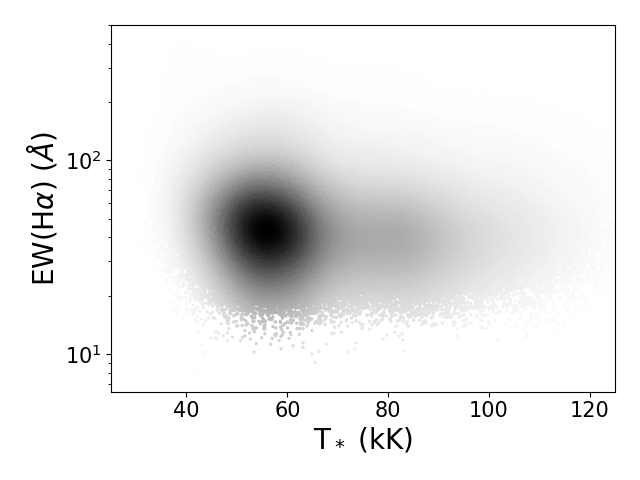}

\caption{Relation between $T_*$ as derived from {\sc HCm-Teff}  and the equivalent width of H$\alpha$ in \AA\ in the MaNGA sample}

              \label{teff-ewha}%
    \end{figure}

Given the results obtained by {\sc HCm-Teff} using these specific emission lines, we can explore to what extent it is reasonable to expect that a large fraction of the regions observed in MaNGA have $T_*$ values dominated by a predominant population of HOLMES. This can be done by deciphering
whether or not there is any correlation between the derived $T_*$ values and other observational spectral features indicative of the average evolutionary status of the ionizing stellar population. 
For instance, the relation between the derived $T_*$ values  and the equivalent width of \ha\ measured within the 
region apertures in the MaNGA survey selected for this work does not show a clear correlation (i.e., the Spearman correlation coefficient, $\rho_S$ is -0.04), as can be seen in Figure \ref{teff-ewha}.
Moreover, regions with a derived $T_*$ $>$ 60 kK do not present a noticeably lower EW(\ha) on average
(i.e., 79 \AA) than those with a lower $T_*$ value (85 \AA).
 
Therefore, even taking into account the fact that the continuum signal has a large contribution from the underlying disk stellar population in the IFS data, which can contain HOLMES (e.g., \citealt{zhang17}), no evidence is found to support the hypothesis that certain regions have larger $T_*$  values due to the increased presence of more evolved, harder stellar populations compared to young ionizing ones.

\subsection{The softness diagram based on [\oii]/[\oiii] and [\sii]/[\ariii]}

\begin{figure}
   \centering
\includegraphics[width=0.4\textwidth,clip=]{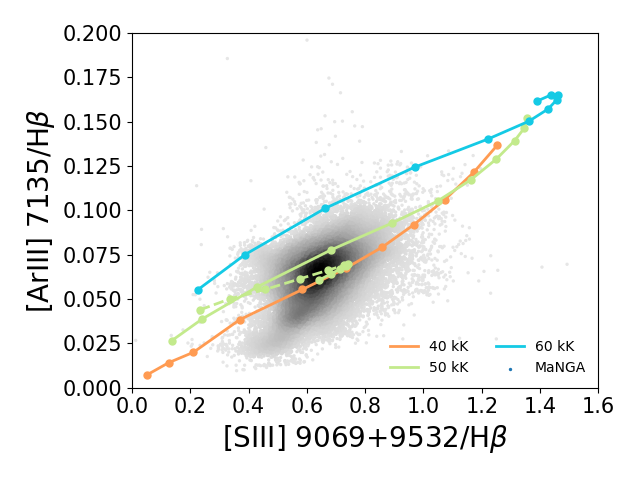}

\caption{Relation between the reddening-corrected relative-to-{\hb} flux of [\siii]  $\lambda\lambda$ 9069+9532 \AA\ and [\ariii] $\lambda$ 7135 \AA\ in the MaNGA sample.
Some model sequences for different values of $T_*$ (40 kK in red, 50 kK in green, and 60 kK in blue) and 12+log(O/H) = 8.9 (solid lines) and 8.0 (dashed line) are included. }

              \label{s3-ar3}%
    \end{figure}

\begin{figure*}
   \centering
\includegraphics[width=0.45\textwidth,clip=]{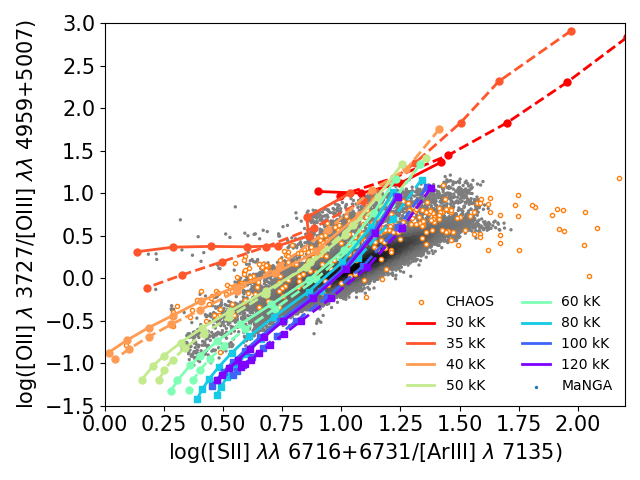}
\includegraphics[width=0.45\textwidth,clip=]{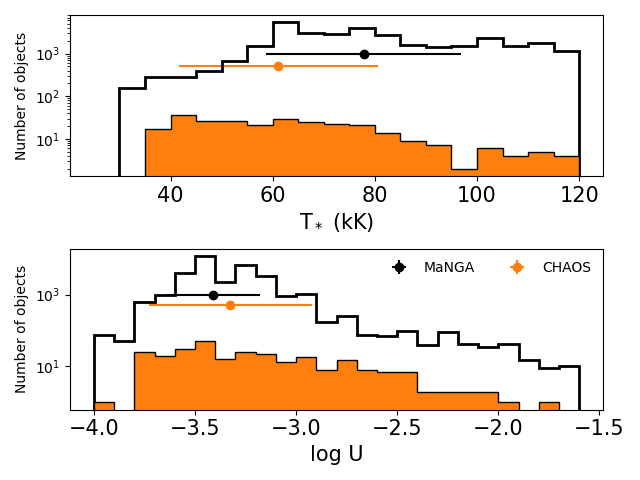}

\caption{Same as figure \ref{o2o3-s2s3} but for [\oii]/[\oiii] vs [\sii]/[\ariii], but only for models with spherical geometry.}

              \label{o2o3-s2ar3}%
    \end{figure*}

Among the possible sources of uncertainty in the results obtained using the  softness diagram,
we can explore the use of [\siii] lines, whose atomic coefficients
have traditionally shown some known discrepancies with results from photoionization models (e.g., \citealt{diaz91,badnell15}). 
In addition, [\siii] line fluxes can have large  uncertainties due to the presence of absorption bands in the
spectral range where the lines at $\lambda$ 9069 and 9532 \AA\ are measured \citep{diaz85}.  

As an alternative to [\siii] lines, we can use the [\ariii] $\lambda$ 7135 {\AA} line, given the similarity of the ionization potentials of S$^{+}$ and Ar$^{+}$ and that S and Ar are $\alpha$-elements that are unaffected by dust depletion.  Figure \ref{s3-ar3} shows the relation between the reddening-corrected emission-line fluxes  of [\siii] $\lambda\lambda$ 9069+9532 \AA\ and [\ariii] $\lambda$ 7135 \AA,\
relative to {\hb,} for the selected regions in MaNGA. 
As expected from the fact that these lines trace the same excitation region, the correlation is clear,
although the correlation coefficient is moderate (i.e., $\rho_S$ = 0.49).
The sequences of photoionization models shown in the same panel indicate that the observational scatter is not owing to the $U$ variation
(i.e.,  in each sequence varying from log $U$ = -4.0 to -1.5, $\rho_S$ is always higher than 0.99), but additional dependences on $T_*$ and O/H exist.
In addition, given the larger wavelength baseline of the [\siii] lines in relation to that of their closer H{\sc i} recombination line, the derived flux of the former can be more affected by the uncertainty in the reddening correction or the presence of telluric emission. 
Therefore, we can  replace the [\siii]  lines  in the corresponding  softness diagram and investigate the effect of this replacement on the results.

The left panel of Figure \ref{o2o3-s2ar3} shows the new softness diagram with the emission-line ratio [\sii]/[\siii] replaced with [\sii]/[\ariii].
We first note that the number of regions for which this ratio can be measured with an S/N of  at least 10 is significantly smaller for MaNGA (i.e., 32\,602 regions), given that the [\ariii] line is weaker than [\siii].
In any case, the behavior observed in the previous diagram, namely the large fraction of  regions with $\eta\prime$ values compatible with the ionization from sources hotter than O stars,  is even more pronounced for both samples.

The right panel of Figure \ref{o2o3-s2ar3} shows the corresponding distributions of the resulting $T_*$ and log $U$ from {\sc HCm-Teff} when we use the emission lines shown in the softness diagram of Figure \ref{o2o3-s2ar3}
 as input (i.e., [\oii], [\oiii], [\sii] and [\ariii]).
The results agree with what can be seen in the softness diagram, as larger fractions of the regions both in MaNGA and CHAOS present $T_*$ larger than 60 kK (i.e., 90\% and 54\%, respectively). The corresponding median values of the
derived $T_*$  also significantly increase
for both samples in relation to the previous case, with values of 78 $\pm$
12 kK for MaNGA and 61 $\pm$ 12 kK for CHAOS, though they remain compatible within the errors.

On the other hand, contrary to $T_*$, the differences in the resulting log $U$ distribution are smaller, even though the mean log $U$ in MaNGA (-3.41 $\pm$ 0.10) is still lower
than in CHAOS (i.e., -3.31 $\pm$ 0.21), albeit compatible within the errors. 
These numbers indicate, on one hand, that the two samples still behave in a different manner when this set of lines is used to derive both $T_*$ and $U$. 
On the other hand, as opposed to the results obtained when the [\siii] lines are used, the results from this softness diagram ---which implies an even larger fraction of regions that are supposedly ionized by HOLMES---  cannot be interpreted as due to the use of the [\siii] emission line.

\subsection{Replacing [\sii] with [\nii] in the softness diagrams}

\begin{figure*}
   \centering
\includegraphics[width=0.45\textwidth,clip=]{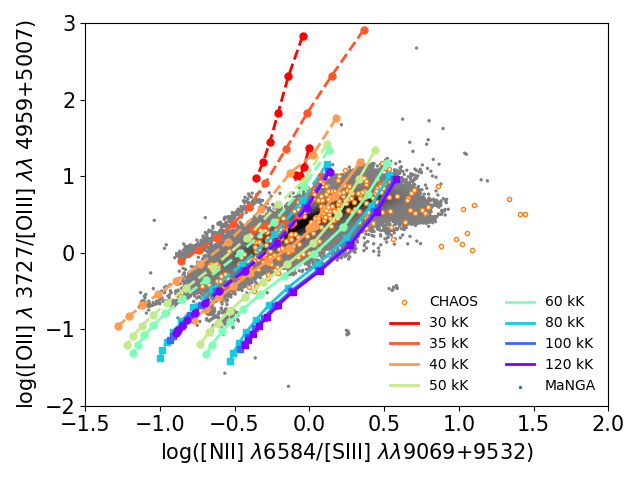}
\includegraphics[width=0.45\textwidth,clip=]{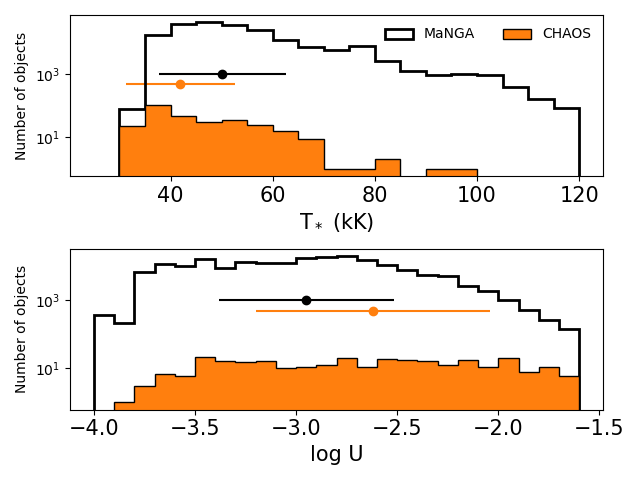}

\caption{Same as figure \ref{o2o3-s2s3} but for [\oii]/[\oiii] vs [\nii]/[\siii] and only for models with spherical geometry.}

              \label{o2o3-n2s3}%
    \end{figure*}

Another factor that may account for the very high values of the emission-line ratio [\sii]/[\siii] in the softness diagram is the background DIG contribution to the
       integrated star-forming region fluxes (e.g., \citealt{zurita2000,haffner09}). 
Given the larger spatial aperture in surveys like MaNGA, which is based on IFS, as compared
with CHAOS, this contribution is expected to be greater. 
Indeed, [\sii] emission (relative to \ha) is known to be enhanced in the DIG with respect to the
         \hii\ regions \citep{reynolds85, dm94, galarza99}.

Therefore, we investigated the role of this emission line in the softness diagram by replacing this line with another strong low-excitation line, namely [\nii] $\lambda$ 6583 \AA. This line can also be contaminated by DIG emission, although to a lesser extent because of the higher ionization potential of N$^+$ than that of $S^+$ (e.g., \citealt{blanc09}), and can also be used to trace both the excitation and the shape of the SED in this diagram.

The left panel of Figure \ref{o2o3-n2s3} shows this new diagram, where [\sii] in replaced with [\nii] both in the models and the data. Contrary to the previous case with [\ariii], the number of regions
in MaNGA for which we can build this diagram is again higher, as this line is bright and easily measurable in most of the regions.
As can also be seen, contrary to the previously shown diagrams based on the [\sii] lines, no clear difference is observed between the two samples and, in addition, only a small fraction of them
apparently lie in the region of the PNe models (i.e., with $T_*$ $>$ 60 kK, supposedly ionized by HOLMES). 
Although the model sequences for different values of $T_*$  apparently present a larger dependence on metallicity (and therefore on N/O) in this new diagram, the fraction of regions with $T_*$ $>$ 60 kK for both samples does not present a mean total oxygen abundance that is larger within the errors (i.e., 12+log(O/H) = 8.58) than the rest of regions,
and, in any case, the O/H values of all the regions are well covered by the grid of models. 

As in previous subsections, in the right panel of Figure \ref{o2o3-n2s3} we show the distributions of $T_*$ and log U calculated by {\sc HCm-Teff} using only the corresponding metallicities and the emission lines involved in the softness diagram of Figure \ref{o2o3-n2s3} as input (i.e., [\oii], [\oiii], [\nii] and [\siii]). In this case, the replacement of [\sii] with [\nii] leads to a substantial difference in the results,
as the number of regions with a resulting $T_*$ $>$ 60 kK is drastically reduced  in both samples (i.e., 20\% in MaNGA and 10\% in CHAOS).
Consistently, the median $T_*$ derived from {\sc HCm-Teff} in MaNGA (50 $\pm$ 11 kK) and CHAOS (42 $\pm$ 8kK) are much more similar, both lying in the regime expected for massive young stars.

Regarding log $U$, we also find differences in relation to the diagrams based on [\sii], as both  median values in MaNGA (-2.95 $\pm$ 0.31) and  in CHAOS (-2.62 $\pm$ 0.37) 
are significantly higher than in previous cases.
Although the fraction of regions with $T_*$ values that are consistent with ionization from objects harder than massive stars is much lower when using [\nii], it is pertinent to question the extent to which the
fractions of star-forming regions in both MaNGA and CHAOS with an effective $T_*$ $>$ 60 kK can be genuinely ionized by a predominant population of HOLMES in the studied regions.
In this way, taking EW(\ha)  again as a proxy for the mean age of the underlying stellar population, its average is almost identical in the total sample (83 \AA) to that in
those regions with $T_*$ $>$ 60 kK (i.e., 80 \AA) when [\nii] is used, and so the very high $T_*$ found in this subsample does not seem to be caused by
a real ionization from HOLMES in these regions. 
On the other hand, given that   [\nii] can also be contaminated by background DIG emission (e.g., \citealt{poetrodjojo19}), and the higher $T_*$ and lower $U$ values still found in MaNGA in relation to CHAOS, a certain DIG contribution could be responsible for these fractions,
even using the [\nii] line.

In any case, these results may indicate ---assuming that [\sii] emission is more contaminated by DIG than [\nii]--- that a large fraction of
the regions in MaNGA is affected by this contribution from the background.
This does not exclude HOLMES as contributors to the ionization of the background DIG emission included in the MaNGa spaxels,
         but would indicate that HOLMES are not the dominant source of ionization in the regions of the MaNGA sample, as suggested
         by previous work (e.g., \citealt{kumari21}). Therefore, the derived $T_*$ values obtained using  certain low-excitation emission lines in low-resolution IFS surveys      are possibly not representative of the \hii\ regions covered by large apertures.

This result is consistent with the very different N/O ratios derived in MaNGA by \cite{z21} when
using the N2O2 or the N2S2 emission-line ratios, which are much lower in the second case, as a consequence of a possible stronger DIG contamination of [\sii] of the \hii\ region fluxes.

\section{Summary and conclusions}

Using different versions of the softness diagram, involving different ratios of low- to high-excitation emission lines, we investigated the different behavior in this diagram of the observational samples MaNGA and CHAOS.
The direct comparison of the observations in the diagram of [\oii]/[\oiii] versus [\sii]/[\siii]   with sequences of photoionization models
of single O stars and PNe reveals that most of the regions in MaNGA have $T_*$ $>$ 60 kK, which is not true for
CHAOS.
As this is the maximum considered $T_*$ value in the models for massive stars, this could be indicative that the MaNGA regions are predominantly ionized by HOLMES, as suggested by \cite{kumari21}.
This is in agreement  with the results from the code {\sc HCm-Teff}; when this code translates the positions in this latter  diagram into equivalent values for $T_*$ and log $U$, the corresponding average values and
the fraction of star-forming regions with $T_*$ larger than the maximum O star value are found to be much higher for MaNGA than for CHAOS.

This result cannot be associated with any problem related to the [\siii] lines at $\lambda\lambda$ 9069,9532 \AA\ (i.e., atomic coefficients, continuum absorption, or telluric contamination), 
as it is reproduced when these lines are replaced with [\ariii] at $\lambda$ 7135 \AA, which traces a similar excitation zone.

However, when [\sii] is replaced with [\nii] in the softness diagram, which is less affected by background DIG, both the median obtained $T_*$ and the fraction of star-forming regions with a value larger than 60 kK are significantly reduced in both samples.

Our results suggest that the DIG contamination may significantly affect the results from this diagram when it is built using [\sii], and that local HOLMES 
are not the predominant source shaping the hardness of the ionizing radiation in the star-forming regions, even in those observed with low spatial resolution, such as in MaNGA.
In any case, the true nature of DIG and its contribution to the integrated spectra are still far from clear, as it is also produced by HOLMES. Also, the lack of correlation between the derived $T_*$ with EW(H$\alpha$), which also depends
on DIG emission (e.g., \citealt{lacerda18}), may indicate that the artificially increased $T_*$ values obtained when certain low-excitation lines are used in low-spatial-resolution data are the product of  a combination of emission lines with a different spatial origin rather than the consequence of a characteristic ionizing stellar population.

\begin{acknowledgements}
This work has been partly funded by projects Estallidos7 PID2019-107408GB-C44
(Spanish Ministerio de Ciencia e Innovacion),
and the Junta de Andaluc\'\i a for grant P18-FR-2664.
We also acknowledge financial support from the State Agency for Research of the Spanish MCIU through the "Center of Excellence Severo Ochoa" award to the Instituto de Astrof\'\i sica de Andaluc\'\i a  (SEV-2017-0709).
AZ and EF acknowledge support from projects PID2020-114414GB-100 and PID2020-113689GB-I00 financed by MCIN/AEI/10.13039/501100011033, from projects P20-00334 and FQM108, financed by the Junta de Andaluc\'i a, and from FEDER/Junta de Andaluc\'i a-Consejer\'i a de Transformaci\'on Econ\'omica, Industria, Conocimiento y Universidades/Proyecto A-FQM-510-UGR20.
EPM also acknowledges  the assistance from his guide dog Rocko without whose daily help this work would have been much more difficult.

\end{acknowledgements}

%
   \bibliographystyle{aa} 
   \bibliography{refs.bib} 
%

\end{document}